\documentclass{JHEP3}

  \usepackage{latexsym,bm,amsmath,amssymb,amsfonts}
  \usepackage{epsfig,graphics,graphicx,mathrsfs}
  \usepackage{slashed}
  \usepackage{cite}
  \usepackage[latin1]{inputenc}

  \long\def\comment#1{ }

  \newcommand{\mcal}{\mathcal}
  
  \newcommand{\beq}{\begin{eqnarray}}
  \newcommand{\eeq}{\end{eqnarray}}
  
 \def\simge{\mathrel{%
   \rlap{\raise 0.511ex \hbox{$>$}}{\lower 0.511ex \hbox{$\sim$}}}}
\def\simle{\mathrel{
   \rlap{\raise 0.511ex \hbox{$<$}}{\lower 0.511ex \hbox{$\sim$}}}}

\vspace{1.5cm}

\preprint{
  OIQP 09-13}

\title{\rm \LARGE Odderon  in baryon--baryon scattering from the AdS/CFT correspondence}

\author{Emil Avsar\\ 104 Davey Lab, Penn State University, University Park, 16802 PA,  USA\\
 E-mail: \email {eavsar@phys.psu.edu}}

\author{Yoshitaka Hatta\\Graduate School of Pure and Applied Sciences, University
of Tsukuba, Tsukuba, Ibaraki 305-8571, Japan\\
E-mail: \email
{hatta@het.ph.tsukuba.ac.jp}}

\author{Toshihiro Matsuo\\Okayama Institute for Quantum Physics,  Kyoyama 1--9--1, Okayama 700--0015, Japan\\
E-mail: \email
{tmatsuo@yukawa.kyoto-u.ac.jp}}

\abstract{
 Based on the AdS/CFT correspondence, we present a holographic description of  
various $C$--odd 
 exchanges in high energy baryon--(anti)baryon scattering, 
  and calculate their respective contributions to the difference in the total cross sections. 
We show that, due to the warp factor of $AdS_5$,  the single Odderon exchange gives a larger total cross section in baryon--baryon collisions than in baryon--antibaryon collisions at asymptotically high energies.
}

\begin{document}
\section{Introduction}

The Odderon is a Regge exchange carrying odd charge parity, $C = -1$ \cite{Lukaszuk:1973nt,Ewerz:2003xi}, 
as compared to its more well--known $C=+1$ partner, the Pomeron. It is a hypothetical excitation lying 
on the $C$--odd glueball Regge trajectory (meaning that the 
relevant amplitudes go like $s^{\alpha(t)}$ in the Regge limit $s\gg |t|$) with  a putative intercept 
$\alpha(0)\sim 1$, and is usually considered 
as a separate entity from mesonic $C$--odd states like the vector meson (`Reggeon') trajectory.  
 While the Pomeron is responsible for the observed rise of 
the $pp$ total cross section, the Odderon should manifest itself in the 
differences between $pp$ and $p\bar{p}$ cross sections, as well as in other exclusive reactions which involve 
 $C$--odd exchanges \cite{Gauron:1986nk,Bartels:2001hw,Dosch:2002ai,Hagler:2002nh,Bartels:2003zu,Bzdak:2007cz,Avila:2006wy,Pire:2008xe,
Merino:2009nu,Domokos:2009hm}.  However,  it has turned out be rather difficult 
to prove the existence of the Odderon unambiguously, partly because the existing experimental data are not
 sufficient or accurate enough. The bulk of the total cross section difference  in $pp$ and
 $p\bar{p}$ collisions
 \beq
 \Delta \sigma(s) \equiv \sigma^{p\bar{p}}(s)-\sigma^{pp}(s)\,, \label{total}
 \eeq
  appears to be well described by the Reggeon exchange with an intercept 
$\alpha_R(0)\sim 0.5$ (namely, $\Delta\sigma \sim s^{\alpha_R(0)-1}$), leaving little room for the Odderon 
contribution.
Other signatures are inconclusive as well, except perhaps some positive indication from the differential 
elastic cross section $d\sigma/dt$ for $pp$ and $p\bar{p}$ collisions performed at the CERN
ISR. [See, \cite{Ewerz:2003xi} for a comprehensive review.]


Describing the observed cross sections for $pp$ and $p\bar{p}$ collisions is 
not feasible using perturbative QCD (pQCD) alone, as the dominant, soft processes
involve large values of the coupling constant. One is therefore led to use  
models  in order to fit the data, and indeed, 
for the Pomeron there exists a very rich and broad phenomenological analysis
 \cite{Donnachie:1992ny}. 
On the other hand, the Odderon accounts for a small fraction of the total cross section, and therefore 
it is extremely sensitive to the details of various assumptions in the fitting parameterizations.  
 Clearly, guidance from first principle calculations is highly desirable in order to better constrain  
and discriminate different models.

One possible way to approach the nonperturbative, strong coupling regime of gauge theories is  the AdS/CFT
 correspondence which relates weakly 
coupled type IIB superstring theory on $AdS_5 \times S^5$ to strongly 
coupled $\mathcal{N}=4$ supersymmetric Yang--Mills (SYM) theory. Of course, 
the SYM theory, being maximally supersymmetric, conformal and having non-chiral 
particle multiplets, is 
in many aspects quite different from QCD. However, given the fact that we 
almost completely 
lack any insight into the strongly coupled dynamics of QCD  from first principles, 
one can study the problem in this different setting and try to extract universal features  that QCD might possibly share.

In this paper, based on the AdS/CFT
correspondence, we propose a coherent description of various $C$--odd exchanges in ${\mathcal N}=4$  SYM 
and calculate their respective contributions to   
high energy baryon--baryon and  baryon--antibaryon scattering. 
 Previously,  in \cite{Brower:2008cy} the Odderon was identified on the string theory side with the 
fluctuations of the antisymmetric Kalb--Ramond two--form, $B$, and its Regge intercept  at 
strong coupling was derived.
Here we extend the work of \cite{Brower:2008cy} in several important directions.   
First, we point out that the vector meson (Reggeon) exchange can be naturally accommodated in the
strong coupling formalism
as  different tensor components of the B--field which were not considered in  \cite{Brower:2008cy}, thereby 
suggesting that the Odderon and the Reggeon are unified in ten--dimensions. This interpretation naturally 
comes from an inspection of the corresponding ${\mathcal N}=4$ SYM operator, and we shall determine its
 anomalous dimension as well as the Reggeon intercept. 
 Next we  study the coupling of the different components of the B--field to ``baryons'' 
which are on the string theory side represented by wrapped D--branes. 
As we shall emphasize 
throughout this paper, the discussion of the Odderon would never be complete without knowing how it couples to 
external objects. This is  in fact  a subtle but crucial problem already in weak coupling perturbation 
theory, and even more so in the AdS/CFT context.   

Our main calculations are performed in Section \ref{baroddcoup}. We evaluate the forward 
baryon--(anti)baryon scattering amplitude from the single Odderon exchange contribution using the exact B--field 
propagator, and  extract the leading contribution to the imaginary part which is related to the 
total cross section difference. Interestingly, we shall find that the
 Odderon exchange gives a {\it larger} cross section in baryon--baryon scattering than in
 baryon--antibaryon scattering, namely, $\Delta \sigma<0$ at very high energies.   
This looks counterintuitive based on our experience with ordinary  flat space calculations---the exchange 
of fields described by differential forms such as the B--field would imply $\Delta\sigma>0$, and this 
is indeed the case for the Reggeon exchange. As we shall see, the extra overall minus sign in the 
Odderon case is essentially due to the warp factor of $AdS_5$, and therefore represents one of the 
hallmarks of gauge/gravity duality.  

The emerging picture from our analysis is that the total cross section difference 
(\ref{total}) has two components; the Odderon which tends to give 
$\Delta \sigma < 0$, and the Reggeon which tends to give $\Delta\sigma>0$. 
 In the final section, we shall interpret the existing experimental data in view of our findings and 
discuss the possibility to observe a negative cross section difference  
 $\Delta\sigma<0$ in QCD at very large $s$.
Actually, the possibility of having $\Delta\sigma<0$ at large energies was discussed 
already in the very first paper on the Odderon \cite{Lukaszuk:1973nt}, and 
more recently in \cite{Avila:2006wy} on the basis of an extrapolation of phenomenological fits 
to the LHC energy scale.

\section{Odderon at weak and strong coupling}
\label{od}

We will start this section by giving a short overview of the status 
of the Odderon within pQCD. As mentioned already, one of the important aspects of the Odderon 
is its coupling to external objects, so we recall how this issue arises in the weakly coupled problem. 
Then in section \ref{oddsym} we discuss the Odderon 
in strongly coupled $\mcal{N} = 4$ SYM. We first describe the argument behind the 
identification of the B--field with the Odderon, and then  
list the Regge intercepts for all the relevant components of the ten--dimensional B--field. 
Then in section \ref{oddoperator} we discuss their physical interpretations in terms of the 
gauge invariant dual operators in the $\mcal{N} = 4$ theory.

\subsection{Odderon in weakly coupled QCD}
\label{oddweak}

The existence of the Odderon is predicted by pQCD. To lowest order, it is 
given by a symmetric color singlet $C$--odd 
combination of three gluons in the $t$--channel. Higher order corrections containing logarithms
of energy $(\ln s)^n$ can be resummed 
by the Bartels--Kwiecinski--Praszalowicz (BKP) equation \cite{Bartels:1980pe,Kwiecinski:1980wb} 
which describes the pairwise interaction of three Reggeized gluons. 

The BKP equation has attracted much attention because of its 
connection to integrability. The equation was originally derived in momentum space, 
and it was later understood that its Fourier transformed formulation in 
coordinate (impact parameter) space is, under certain assumptions, 
identical to the eigenvalue problem of an exactly solvable spin--chain model (namely the 
XXX Heisenberg spin $s=0$ model) \cite{Lipatov:1993yb,Faddeev:1994zg}. So far, two exact solutions of 
the BKP equation have been found; the Janik--Wosiek (JW) solution  \cite{Janik:1998xj} (see, also, 
\cite{Kotanski:2006ec}) with a Regge intercept\footnote{For a reason to become 
clear later, henceforth we distinguish the intercept associated with the spin  of gauge theory operators $j_O$ from the 
one associated with the total cross section $\alpha_O(0)$.}
slightly less than 1, $j_O(0) \approx  1-0.2472
\frac{\alpha_s N_c}{\pi}$, and the Bartels--Lipatov--Vacca (BLV) solution \cite{Bartels:1999yt} with 
    $j_O=1$.  
   
   Apart from the  difference in the intercept, these two solutions have markedly different coordinate 
dependences which crucially determine their relevance in physical amplitudes. The JW solution is 
constrained such that it vanishes when two of the three gluons sit at the same point. This means that 
it does not couple to $q\bar{q}$ states including a virtual  photon in DIS.  Though it in principle 
couples to three--quark states (like a proton),  the  constraint of vanishing at equal points is 
incompatible with gauge invariant initial conditions \cite{Hatta:2005as}.     
   
   On the other hand, the BLV solution does not have this constraint, and hence it  naturally couples 
to both $q\bar{q}$ and $qqq$ states in a gauge invariant way \cite{Kovchegov:2003dm,Hatta:2005as}.
 Nevertheless, this solution is sometimes deemed `exceptional' because its description in the
 integrability framework turned out to be tricky: One has to extend the Hilbert space of 
eigenstates of the spin--chain Hamiltonian so that it includes solutions which do not 
vanish at equal points.  However, 
physically there is no reason to expect that the Odderon amplitude vanishes at equal points, 
and indeed the BKP equation does not {\it a priori} imply such a property. In fact, this constraint 
was artificially imposed by hand by neglecting  certain terms in the full BKP Hamiltonian in order
 to establish the equivalence with the spin--chain Hamiltonian. Therefore, the seemingly unusual 
status of the BLV solution is illusory, and one can fairly conclude that it is {\it the} Odderon 
solution in perturbative QCD in the leading logarithmic approximation. Note that, as was 
originally done in \cite{Bartels:1999yt}, the BLV solution can be  explicitly constructed in 
momentum space without any reference to integrability. 
  
  The above situation in perturbative QCD illustrates an important point when studying the 
Odderon: It is one thing to  derive the Odderon intercept, but it is quite another to discuss 
{\it how} a given Odderon solution actually couples to external states.  
Whether it has nonvanishing coupling depends on the quantum numbers and the internal 
structure of the colliding hadrons.   As we shall see, this problem appears also in 
the approach based on the AdS/CFT correspondence.

  Another point worth noting is that the BLV odderon amplitude is purely real. This is most readily
 seen by the identification of the odderon amplitude as the imaginary part of the eikonal amplitude 
made up of lightlike Wilson lines \cite{Hatta:2005as}  
  \beq
  {\mathcal A}_{odderon} \sim {\rm Im}\, {\rm tr}\, W\,.
  \eeq
   This implies, in particular, that in perturbation theory one cannot really address the issue of 
the total cross section difference which, via the optical theorem, is related to the imaginary part of 
the forward amplitude ${\rm Im}\, {\mathcal A}(t=0)$.  
   In order to generate a nonzero imaginary part, one has to have some nonperturbative inputs, and
 this is what we shall explore in the following.
   
   Before leaving this subsection, we should like to mention that  there have been a lot of activities
 in phenomenological applications \cite{Bartels:2001hw,Dosch:2002ai,Hagler:2002nh,
Bartels:2003zu,Bzdak:2007cz,Avila:2006wy,Pire:2008xe} 
as well as theoretical extensions of the BLV odderon treatment. 
The latter includes the corrections of next--to--leading logarithmic terms and beyond 
\cite{Braun:2007kz, Stasto:2009bc}, suggesting that the Odderon intercept will remain at $j_O(0)=1$
to all orders in (leading--twist) perturbation theory, and also non--linear effects from 
gluon saturation \cite{Kovchegov:2003dm,Braunewell:2005ct,Hatta:2005as,Kovner:2005qj,Motyka:2005ep,
Jeon:2005cf} which tend to suppress the odderon amplitude.

 \subsection{Odderon in strongly coupled ${\mathcal N}=4$ SYM}
\label{oddsym}

In this and the next subsections we describe the nature of the Odderon in 
${\mathcal N}=4$ supersymmetric
 Yang--Mills (SYM) theory at strong coupling. Via the AdS/CFT correspondence, the problem can be
 equivalently formulated in type--IIB superstring theory on the background of $AdS_5\times S^5$ with the metric 
\beq
ds^2 &=& G_{mn}dX^m dX^n + G_{\alpha\beta}d\theta^\alpha d\theta^\beta \nonumber \\
&=& R^2\frac{-2dx^+dx^-  + d\vec{x}_\perp^2+dz^2}{z^2}+R^2(d\theta^2 + \sin^2 \theta d\Omega_4^2)\,,
\eeq
 where  $x^\pm = \frac{1}{\sqrt{2}}(x^0 \pm x^3)$, $\vec{x}_\perp=(x^1,x^2)$ and $R$ is the common radius of $AdS_5$ and $S^5$. 
We use the notation $X^m=(x^\mu,z)$ for the $AdS_5$ 
coordinates and  $\theta^\alpha=(\theta\equiv \theta^1,\theta^2,\theta^3,\theta^4,\theta^5)$ for the 
$S^5$ coordinates.

In $\mathcal{N}=4$ SYM where the fermions belong to a real (namely, the adjoint) representation of the gauge group, $C$--conjugation is a rather uninteresting operation.   However, if one introduces charges (`quarks') in the fundamental representation living on some `flavor' branes, one can define the $C$--conjugation with respect to these charges. Specifically, 
we shall later introduce baryons, or D--branes which carry the NS-NS charges on their worldvolume.
 The NS-NS antisymmetric B--field couples to the   fundamental and anti-fundamental charges with 
opposite signs, thus it can be identified as the Odderon in this context.\footnote{If one is interested 
in colliding objects which carry the R-R charges,  such as D1--branes and monopoles, the R-R two--form would serve as the Odderon.} 

More properly,  the Odderon is the {\it Reggeized} B--field,\footnote{The exchange of the bare (i.e., not Reggeized) B--field in high energy scattering 
was previously considered in \cite{Janik:1999zk}.  } which roughly means that it is a coherent superposition of string excited states lying on a Regge trajectory starting from the massless B--field. 
 In flat ten--dimensional spacetime, this can be seen by inspection 
of the Shapiro--Virasoro amplitude and takes the form
\beq
f(\alpha' t)(1-e^{-i\pi \alpha(t)}) s^{\alpha(t)}\,, \label{amp}
\eeq
where $\alpha(t)=1+\frac{\alpha't}{2}$. One recognizes the usual Regge behavior with the negative signature factor,
 but importantly the prefactor $f(\alpha' t)$ has no pole at $t=0$. This means that the massless B--field 
decouples from the $s$--channel closed strings such as the dilaton (`glueball'). 
The fact that one nevertheless gets a nonzero amplitude with an imaginary part (\ref{amp}) is due to the Reggeization of the B--field. 

The analogous problem in the background of $AdS_5\times S^5$ is considerably more complicated. Upon compactification on $S^5$, 
the ten--dimensional B--field splits into the diagonal modes
\beq
B_{mn}\,, \label{diag}
\eeq
with both indices along the $AdS_5$ direction, and the off--diagonal modes
\beq
B_{m\alpha}\,, \label{offdiag}
\eeq
   with one of the indices along the $S^5$ direction.\footnote{We ignore the modes $B_{\alpha\beta}$ 
with both indices on $S^5$ because they are irrelevant at high energy.} Furthermore, each mode undergoes
 the Kaluza--Klein (KK) decomposition  \cite{Kim:1985ez}
\beq
B_{mn}&=&\sum_{k=0}^\infty B^{(k)}_{mn}(X) Y^{(k)}(\Omega_5)\,,\nonumber \\
 B_{m\alpha}&=&\sum_{k=1}^\infty \left( b^{(k)}_{m}(X) Y_\alpha^{(k)}(\Omega_5)+ d^{(k)}_m(X) 
\nabla_\alpha Y^{(k)}(\Omega_5) \right)\,, \label{ex}
\eeq
where $Y^{(k)}$ and $Y_\alpha^{(k)}$ are the rank--$k$ scalar and vector spherical harmonics 
on $S^5$, respectively. These harmonics are reviewed  in Appendix A. 
Summation over different harmonics with the same value of $k$ is 
understood. By choosing the gauge $\nabla^\alpha B_{m \alpha}=0$, one can always 
set $d^{(k)}_m=0$ (see (A.14) ) which we shall subsequently do. 

The Reggeization of the B--field has to be done for each KK mode by analyzing the supergravity 
equation of motion \cite{Brower:2006ea}. For the diagonal modes (\ref{diag}), there are two equations 
of motions for each value of $k$. Accordingly, there are two branches of the Regge intercepts.  They
have been worked out  in \cite{Brower:2008cy}  with the results
   \beq
   j_O=1-\frac{M^2_I}{2\sqrt{\lambda}}\,, \qquad (I=1,2) \label{s1}
  \eeq
  where $\lambda$ is the 't Hooft coupling and the respective masses $M_I$ are given by
  \beq
  M_1=k\,, \qquad M_2=k+4\,. \qquad (k=0,1,2.\cdots) \label{kkk}
  \eeq
  Note that the lowest KK state $M_1=k=0$ has $j_O=1$. In view of the finding of  \cite{Stasto:2009bc},
 it is tempting to regard this mode as the fate of the BLV odderon at strong coupling. However, in 
the gravity description this mode is pure gauge, and decouples from physical amplitudes. We shall have
 more to say about this below.
  
The intercept for the off--diagonal mode $B_{m\alpha}\sim b_m$ can be determined as follows.    
   The relevant equation of motion is the massive Maxwell equation  \cite{Kim:1985ez}
   \beq
\nabla^n f_{nm}-\frac{M^2}{R^2} b_m=0\,, 
\eeq
where $f_{mn}\equiv \partial_m b_n-\partial_n b_m$ and 
\beq
M^2=(k+1)(k+3)\,. \qquad (k=1,2,3,\cdots)
\eeq
 Without the mass term, this is identical to the equation for the gauge boson (dual to the
 ${\mathcal R}$--current operator) considered in \cite{Hatta:2009ra}. Thus the intercept can be 
immediately obtained by making the following change in the result of  \cite{Hatta:2009ra}
\beq
j=1-\frac{1}{2\sqrt{\lambda}}\quad \to \quad j_R=1-\frac{1+M^2}{2\sqrt{\lambda}}
=1-\frac{(k+2)^2}{2\sqrt{\lambda}}\,,\label{ther}  \label{s2}
\eeq
 where the meaning of the subscript $R$ will be explained shortly. 
   Interestingly, despite the difference in the equations of motion in $AdS_5$, the spectrums 
of $j$ in (\ref{s1}) and (\ref{s2}) are overlapping except for the first few KK states. 
This might be due to the fact that these modes originally come from a single field, namely the 
B--field, in ten dimensions.

   \subsection{Correspondence with operators in gauge theory}
\label{oddoperator}

The AdS/CFT correspondence identifies fields on the string theory side with local 
gauge invariant operators on the field theory side. Now that we know the intercepts of various 
KK excitations  of the B--field, let us discuss to which operators in ${\mathcal N}=4$ 
SYM  these states correspond. 

Mathematically, the Reggeization means that the spin of the B--field is 
analytically continued to $j\neq 1$. Accordingly, the spin of the corresponding operator should 
also be  continued.\footnote{This implies that the operators become nonlocal.}    
 In the case of the diagonal modes (\ref{diag}), the relation between the dimension of the operator
 with spin $j$ and the mass of the KK state was derived in  \cite{Brower:2008cy}
 \beq
\Delta(j)=2+2\sqrt{\frac{\sqrt{\lambda}}{2}\bigl( j-j_O\bigr)}=2+
2\sqrt{\frac{\sqrt{\lambda}}{2}\left(j-1+\frac{M^2_I}{2\sqrt{\lambda}}\right)}\,.
\label{delta1}
\eeq
Consider first  the branch $M_1=k$ which has a larger intercept and should 
therefore dominate the cross section. One has
\beq
\Delta(1)=2+k\,.
\eeq
As already mentioned, the $k=0$ mode is  pure gauge and therefore does not 
correspond to propagating physical degrees of freedom. Indeed 
there is no gauge--invariant, spin--1, dimension--2 operator in ${\mathcal N}=4$ SYM. 
For $k=1$, the corresponding gauge theory operator is 
\beq
\mbox{tr} \bigl(\psi^A\sigma_{\mu\nu}\psi^B + 2i \phi^{AB}
F^+_{\mu\nu}\bigr)\,, \quad (1\le A,B \le 4)
\label{SYMoperator1}
\eeq
 which belongs to the \textbf{6} representation of $SU(4)$ and the (1,0) of the Lorentz $SO(3,1)$ group. 
[$F^{\pm}$ denotes the self--dual/anti self--dual part of the field strength, 
belonging to $(1,0)$ and $(0,1)$, and the 
trace is in the fundamental representation.] 
 The operators with $k\ge 2$ are multiplied by higher powers of scalars with appropriate $SU(4)$ 
representations.
These operators are denoted as ${\mathcal O}^{(4)}_k$ in Table 7 of \cite{D'Hoker:2002aw}. 

Next, let us turn to the $M_2=k+4$ branch. In this case $\Delta(1)=k+6$, and 
the corresponding operators are given by
\beq
\mbox{tr}\left(F^+_{\mu\nu}F^2_-\phi^k\right)\,.
\label{SYMoperator2}
\eeq
These operators are  denoted as ${\mathcal O}^{(16)}_k$ in Table 7 of \cite{D'Hoker:2002aw}. 
 Note that for $k=0$ this is a gluonic operator which, at weak coupling, starts out with 
three gluons. Based on this analogy, and also on the fact that the fields on the two branches 
$I=1,2$ can be treated on the same footing in actual calculations,  we shall collectively call 
the diagonal modes $B_{mn}^{(k)}$  the Odderon as in \cite{Brower:2008cy}. 

Finally, for the off--diagonal modes, one has instead \cite{Hatta:2009ra} 
\beq
\Delta(j)=2+2\sqrt{\frac{\sqrt{\lambda}}{2}\bigl( j-j_R\bigr)}=2+2\sqrt{\frac{\sqrt{\lambda}}{2}\left(j-1+\frac{1+M^2}{2\sqrt{\lambda}}\right)}\,.
\label{delta2}
\eeq
Using $M^2=(k+1)(k+3)$, 
one  finds $\Delta(1)=k+4$. The corresponding operators 
are denoted as ${\mathcal O}^{(10)}_{k-1}$ 
in Table 7 of \cite{D'Hoker:2002aw} and read
\beq
\mbox{tr} \left( F^+_{\mu\nu}\bar{\psi}_A\bar{\sigma}^\nu \psi_B \phi^{k-1} \right)\,.
\label{SYMoperator3}
\eeq
The lowest $k=1$ mode is the \textbf{15} of $SU(4)$ and the ($\frac{1}{2},\frac{1}{2})$ of $SO(3,1)$. 
It looks like an interpolating operator of  
vector mesons. Therefore we will refer to the off--diagonal modes $b_m^{(k)}$ as the Reggeon, 
whence the subscript in $j_R$.

Concluding this section, we have identified all the possible $C$--odd exchanges originating from
the ten--dimensional B--field and listed their respective Regge intercepts and the corresponding
operators in SYM. 
Anticipating the later developments, here we note the following two caveats 
when applying the formal results of this section to actual scattering processes: (i) 
Our experience 
with pQCD tells us that 
it may very well be that
 not all  
of the KK modes actually couple to external hadrons. 
(ii) The formulas presented above for the Regge ``intercept'' lead to the following behavior
\beq
{\mathcal A}(s, b)\sim s^{j_O-1}f(s,b)\,,
\eeq
of the scattering amplitude at {\it fixed } impact parameter $b$, with $f$ being 
some function of $b$  and the energy $s$. In order to obtain the total 
cross section difference, one has to integrate Im${\mathcal A}(b)$ over $b$. If 
the function $f$ had no energy dependence, then the cross section would have 
the dependence $s^{\alpha_O(0)-1}$ with $\alpha_O(0)=j_O$. 
However, in certain cases the $b$--integration can modify the 
$s$--dependence so that $\alpha_O(0)\neq j_O$.  In the next section, we shall see that both of these caveats are indeed relevant.

\section{Odderon exchange in baryon--baryon scattering at strong coupling}
\label{baroddcoup}

In this main section we calculate the Odderon exchange contribution to the baryon--(anti)baryon 
scattering amplitude by representing baryons as D--branes. Ideally, the D--brane amplitude should 
be exactly calculated (as can be done in flat space \cite{Bachas:1995kx}) so that various string 
exchanges are automatically included.  However, in a curved background like $AdS_5\times S^5$, such exact results are unavailable, and  
 we must content ourselves with the single Odderon exchange. 
   Actually, in ${\mathcal N}=4$ SYM it is probably possible to go beyond the single Odderon approximation \cite{Brower:2008cy} by including unitarity corrections in the form of the graviton (Pomeron) eikonalization
     \cite{Cornalba:2007zb,Brower:2007qh}.     
   We will comment on the issue of unitarity corrections along with their relevance to QCD in the discussion section.

We first describe the D--brane solution of \cite{Imamura:1998gk,Callan:1998iq} and 
investigate its coupling to the various KK excitations of the B--field. It will turn out 
that the coupling to the Reggeon $b_{m}^{(k)}$ vanishes identically.   
We therefore focus only on the Odderon $B_{mn}^{(k)}$ contribution, 
calculating first the bare propagator in section \ref{bareprop} from which the 
Reggeized propagator is immediately obtained. The latter is then used in 
section \ref{sigmadiff} to calculate the Reggeized amplitude whose 
imaginary part determines the cross section difference. 
Our main result is displayed in equation \eqref{fin} which is an explicit analytical 
formula for the total cross section difference.

\subsection{Baryon--Odderon coupling}
In the context of gauge/string duality, baryons can be realized as  D--branes wrapping on some
 compact manifold \cite{Witten:1998xy}.  In the canonical example of  the ${\mathcal N}=4$ 
SYM/type IIB correspondence, they are D5--branes wrapping on $S^5$. Among the different 
versions of baryon configurations proposed in the literature, we employ here a particular
 BPS solution constructed in \cite{Imamura:1998gk,Callan:1998iq}.  
 
  Consider the embedding of a D5--brane in $AdS_5\times S^5$ using the coordinate mapping 
$\beta(\xi) = (x^\mu(\xi), z(\xi), \theta^\alpha(\xi))$ where $\xi^a=(x^0,\theta^\alpha)$ 
are the coordinates parameterizing the worldvolume of the D5--brane.
 The dynamics of the D5--brane is governed by the Born--Infeld action plus the Chern--Simons term
 \beq
 S=T_5 \int d^6\xi \left\{-\sqrt{-\mbox{det}(\tilde{G}+\tilde{B}+2\pi \alpha' F)} +
 2\pi \alpha' F\wedge c_{(4)} \right\}\,,
 \eeq
 where $\tilde{G}\equiv \beta^*G$, $\tilde{B} \equiv \beta^*B$ and 
$c_{(4)} \equiv \beta^* C_{(4)}$ are the 
pullbacks of the graviton, the B--field and the R--R 4--form on the 
brane, respectively,  and $F$ is the two--form $U(1)$ field strength on the brane. 
The solution \cite{Imamura:1998gk,Callan:1998iq} which respects the BPS condition  has  
  $S^4 \subset S^5$ symmetry and is extended in the $z$--direction along which there is an 
electric flux $F_{0z}=(\partial \theta/\partial z)F_{0\theta}$. [Remember that $\theta\equiv \theta^1$.] The embedding function 
$z=z(\theta)$ is given by  
\beq
z(\theta)=\frac{r_h\sin \theta}{\left[\frac{3}{2}(\theta-\sin \theta \cos \theta)\right]^{\frac{1}{3}}}\,, \label{bps}
\eeq
 where $r_h=z(\theta=0)$ is an arbitrary length scale which we identify with the 
radius of the baryon. As $\theta$ goes to $\pi$, $z(\theta)$ monotonously decreases to zero 
where the boundary of $AdS_5$ is located. In order to obtain a finite baryon mass, one has to put a UV cutoff at small $z$, or equivalently, in $\theta$ near $\pi$. 

We shall be interested in the high energy collision of a baryon and a (anti-)baryon 
in this D--brane representation, exchanging 
the B--field in the $t$--channel. Since the mass of a baryon is of order $N_c$, 
high energy  means that the center--of--mass energy $\sqrt{s}$ is parametrically 
of order $N_c$ up to a boost factor. For a baryon moving in the $\pm x^3$  direction near the 
speed of light $v\approx 1$, it is convenient to take $x^\pm$ (instead of $x^0$) as the `time' coordinate
 on the brane with the constraint $x^\mp \approx 0$. The coupling between the D--brane moving in the 
$+x^3$ direction and the B--field is
\beq
S_{int}&=& \frac{1}{\mbox{Vol}_{S^4}} \int dx^+ d\theta d\Omega_4 \frac{\partial 
{\mathcal L}}{\partial F_{+\theta}}\frac{\tilde{B}_{+\theta}}{2\pi \alpha'}
 \nonumber \\
 &=&\frac{n}{2\pi \alpha'\mbox{Vol}_{S^4} } \int dx^+ d\theta d\Omega_4 \left
(B_{+\theta}+\frac{\partial z}{\partial \theta}B_{+z}\right)\,, \label{couple}
\eeq
 where $\mbox{Vol}_{S^4}=\frac{8\pi^2}{3}$ and $n=\frac{\partial {\mathcal L}}
{\partial F_{+\theta}}$ is an integer which measures the string charge on the 
D--brane. For a baryon (anti--baryon) we take $n=N_c$ ($n=-N_c$). 

The B--fields in (\ref{couple}) are 
decomposed into the KK modes as in (\ref{ex}). Then the issue arises as to whether 
the $d\Omega_4$ integral of the spherical harmonics is nonvanishing. 
  In Appendix A, we show that the integral of the $\theta$--component of the vector 
spherical harmonics over $S^4$ vanishes identically for all $k$,
 \beq
 \int d\Omega_4 \, Y_\theta^{(k)}(\Omega_5)=0\,. \label{prove}
 \eeq
This means that the off--diagonal modes $B^{(k)}_{+\theta}$ (Reggeon) decouple from the 
the baryon under consideration.  
On the other hand, for any $k\ge 0$ at least one component of the scalar harmonics
 $Y^{(k)}$ gives a  nonvanishing value after the integration over $S^4$, so  the diagonal modes $B^{(k)}_{+z}$ (Odderon) do 
couple to our baryon. Therefore in the remainder of this section we only consider the 
diagonal mode $B_{+z}$. We shall return to the relevance of the off--diagonal modes in the discussion section.

The full amplitude in impact parameter space 
for the exchange of the B--field is 
  \beq
 i{\mathcal A^{\pm}}(s,b)&=&\pm i^2\left(\frac{N_c}{2\pi \alpha' \mbox{Vol}_{S^4} }\right)^2 
\sum_{k} \int dx^+ dz d\Omega_4 Y^{(k)}(\Omega) \nonumber \\ && \times
  \int dx'^-dz'd\Omega'_4 Y^{(k)}(\Omega')\langle B^{(k)}_{+z}(x^+,0,x_\perp,z)B^{(k)}_{-'z'}
(0,x'^-,x_\perp',z')\rangle \,, \label{full}
  \eeq
  where the plus (minus) sign corresponds to baryon--baryon (baryon--antibaryon) scattering.
Our next task is now to calculate the propagator $\langle B_{+z} B_{-'z'} \rangle$.

\subsection{Bare B--field propagator}
\label{bareprop}

The equation of motion for the components $B^{(k)}_{mn}$  has been derived in  \cite{Kim:1985ez}
by dimensionally reducing the type IIB supergravity equation of motion. Alternatively, one may
 perform the dimensional reduction directly on the supergravity action taking into account the mixing
 with the Ramond--Ramond (R-R) two--form. This yields, for a given value of $k$, \cite{Arutyunov:1998hf}
\beq
S= -\frac{R^5 \pi^3}{2\kappa^2}\sum_{I=1,2}\int d^5X \sqrt{-G}\Bigl(\frac{i}{2}\epsilon^{mnlpq}(a^{I}_{mn})^*
\partial_l a^{I}_{pq} +M_I  (a^{I}_{mn})^*a^{mn}_{I}\Bigr)\,, \label{baction}
\eeq
 where  we have normalized the spherical harmonics as $\int d\Omega_5 |Y^{(k)}|^2=\pi^3=\mbox{Vol}_{S^5}$. 
The complex fields $a^{1,2}$ are certain linear combinations of the B--field and the R-R 
 two--form \cite{Arutyunov:1998hf}, such that the B--field can be written\footnote{Our normalization 
of the B--field differs from that in  \cite{Arutyunov:1998hf}  by a factor of 2.}
 \beq
 B_{mn}^{(k)}=\frac{1}{\sqrt{2(k+2)}} \left( a^{1}_{mn}+a^{1*}_{mn}+a^{2}_{mn}+a^{2*}_{mn} \right)\,.
 \eeq
 The Kaluza--Klein masses $M_I$ are as in (\ref{kkk}).
 
The propagator of the $a^{(1,2)}$ field satisfies the following equation (up to the prefactor $\kappa^2/R^5\pi^3$)
\beq
 &&\frac{i}{2}\epsilon_{mn}^{\ \ \ \, lpq}\partial_lD_{pq,m'n'}(X,X') + MD_{mn,m'n'}(X,X')
 \nonumber \\
 && \qquad  \qquad  =
-i \frac{\delta^{(5)}(X-X')}{\sqrt{-G}}(G_{mm'}G_{nn'}-G_{mn'}G_{nm'})\,,
\eeq
  where $M$ is either $M_1$ or $M_2$.
The solution was obtained in \cite{Bena:2000fp} and takes the form
\beq
D_{mn,m'n'}&=&T^1_{mn,m'n'}(D(u)+2H(u))+T^2_{mn,m'n'} H'(u) + T^3_{mn,m'n'} K(u) \nonumber 
\\ &=& T^1_{mnm'n'}D(u) - \partial_m V_{n,m'n'} + \partial_n V_{m,m'n'}+  T^3_{mn,m'n'} K(u)\,, \label{v}
\eeq
where
\beq
T^1_{mn,m'n'}&=& R^4(\partial_m \partial_{m'} u \partial_n \partial_{n'} u - \partial_m \partial_{n'} u
\partial_n \partial_{m'} u)\,,\label{11} \\
T^2_{mn, m'n'}&=&R^4(\partial_m \partial_{m'} u \partial_n u \partial_{n'} u - \partial_m \partial_{n'} u
\partial_n u \partial_{m'}u \nonumber \\ 
 && \qquad \qquad - \partial_n \partial_{m'} u \partial_m u \partial_{n'} u + \partial_n \partial_{n'} u  \partial_m u \partial_{m'} u) \,,  \\
T^3_{mn, m'n'}&=&R^5\epsilon_{mn}^{\ \ \  lpq}\partial_l \partial_{m'} u \partial_p \partial_{n'} u \partial_q u\,, \\
 V_{m, m'n'}&=&R^4H(u) (\partial_m \partial_{m'} u \partial_{n'} u - \partial_m \partial_{n'} u \partial_{m'} u )\,,
\eeq
 and $u$ is the chordal distance in $AdS_5$
 \beq
 u=\frac{(z-z')^2+(x_\perp-x_\perp')^2-2(x^+-x'^+)(x^- - x'^-)}{2zz'}\,.
 \eeq

We shall focus on the $(mn,m'n')=(+z,-'z')$ component. One can check that $T^3_{+z,-'z'}=0$, and
\beq
T^1_{+z,-'z'}=\frac{R^4}{z^2z'^2}\left(1+v-\frac{z}{z'} - \frac{z'}{z} \right) = \frac{R^4}{zz'} 
\partial_z \partial_{z'} v\,, \label{first}
\eeq
 where
 \beq
 v=\frac{(z-z')^2+(x_\perp-x'_\perp)^2}{2zz'}\,,
 \eeq
 is the chordal distance in $H_3$. As for terms involving $V$, the first term $-\partial_+ V_{z,-'z'}$ 
can be neglected since we integrate the propagator over $x^+$ (see (\ref{full})), while the other term is given by 
 \beq
 \partial_zV_{+,-'z'}
 = \partial_z \left(\frac{H(u)}{zz'}\partial_{z'} v\right)\,.
 \eeq
 The function $D(u)$ satisfies the following equation
\beq
 \frac{1}{R^2} (z^2\partial_z^2 -3z\partial_z + z^2 \partial^2 + 4-M^2)D(u)=iM\frac{z^5}{R^5}\delta^{(5)}(X-X')\,, \label{from}
 \eeq
  and $H(u)$,  $K(u)$ are given in terms of $D(u)$
  \beq
  H(u)=-\frac{1}{M^2}\left(2D(u) + (u+1)D'(u)\right)\,, \quad K(u)=-\frac{i}{M}D'(u)\,.
  \eeq
Let us define
 \beq
 R^2\int \frac{dx^+dx'^-}{zz'} D(u)=D^{(3)}(v)\,.
  \eeq
  Then
  \beq
   R^2\int \frac{dx^+dx'^-}{zz'} H(u)=-\frac{1}{M^2}D^{(3)}(v)\,,
 \eeq
 where we integrated by parts.
 From (\ref{from}), one has
  \beq
  \frac{1}{R^2}(z^2\partial_z^2 -z\partial_z + z^2 \partial^2_\perp + 1-M^2)D^{(3)}=iM\frac{z^3}{R^3}\delta(z-z') \delta^{(2)}(x_\perp-x'_\perp)\,, \label{right}
  \eeq
   and the solution can be conveniently written as 
         \beq
   D^{(3)}(v)=\frac{-iM}{4\pi R} \frac{e^{-M\xi}}{\sinh \xi}\,, \label{g3}
\eeq
 where  $\xi \equiv \cosh^{-1}(1+v)$. Note that $D^{(3)}$ vanishes when $M=0$, which explicitly shows the decoupling of  the mode $M_1=k=0$. [The $V$ terms become gauge artifacts.] In the following we consider only the case $M\ge 1$ and obtain
\beq
&&\int dx^+ dx'^- \langle B^{(k)}_{+z}(x,z)B^{(k)}_{-z}(x',z')\rangle = \sum_{I}\frac{1}{k+2}\frac{\kappa^2}{R^5\pi^3}
R^2  \nonumber \\
 && \qquad \qquad \qquad \times \left[ \left(1-\frac{1}{M^2_I}\right)D^{(3)}(v)\partial_z \partial_{z'} v
 - \frac{1}{M^2_I}\partial_v D^{(3)}(v) \partial_z v \partial_{z'}v \right]\,.  \label{get}
\eeq

\subsection{The total cross section difference}
\label{sigmadiff}

Equation \eqref{get} is the contribution from the bare B--field, and as such, the corresponding amplitude ${\mathcal A}$  is purely real. In order to 
get an imaginary part, one has to 
 Reggeize the B--field by doing analytic continuation in $j$. 
This boils down to replacing  \cite{Brower:2008cy}\footnote{Note that we do not replace 
$M\to M_j$ in the prefactor. This factor of $M$ comes from the right hand side of (\ref{right}) 
and is clearly not associated with the pole of the $t$--channel propagator. 
Incidentally, Ref.~\cite{Brower:2008cy} leaves open the possibility that the $M=0$ mode becomes physical after the analytic continuation. In our approach this might correspond to replacing $M\to M_j$ also in the prefactor so that the coupling apparently becomes nonvanishing. However, such a procedure is somewhat  arbitrary, and induces  uncertainties in the subsequent calculations. Thus, although we think it is an interesting possibility, we do not pursue it in the present paper. }
\beq
D^{(3)}(v) \to D^{(3)}_j(v)\equiv \frac{-iM}{4\pi R} \frac{e^{-M_j\xi}}{\sinh \xi}\,, \label{rep} 
\eeq
where
\beq
 M_j^2 = M^2 + 2\sqrt{\lambda} (j-1) =2\sqrt{\lambda}(j-j_{O})\,,
\eeq
 and  $j_{O}\equiv 1-\frac{M^2}{2\sqrt{\lambda}}$ is the Odderon intercept (\ref{s1}).
 Let us  consider the first term on the right hand side of (\ref{get}). 
Summing over odd values of $j$ in the form of the contour integral, as dictated by the
presence of the odd signature factor in (\ref{amp}), one reaches the following expression
\beq
\int \frac{dj}{4i}\frac{1-e^{-i\pi j}}{\sin \pi j} \left(\frac{\alpha' \tilde{s}}{4}
\right)^{j-1} \left(1+v-\frac{z}{z'} - \frac{z'}{z} \right)\frac{D^{(3)}_j(v)}{zz'}\,. \label{inter}
\eeq
 The $j$--integral goes
around the positive real axis encircling poles at odd integers $j=1,3,5,\cdots$ clockwise.
In order to obtain the forward scattering amplitude ${\mathcal A}(s,t=0)$ one needs to take the Fourier 
transform of (\ref{inter}) with respect to $b=x_\perp-x'_\perp$ and take the limit  
$t\to 0$.\footnote{It is important to notice that the limit $t\to 0$ has to be taken {\it after} the $j$--integration, which means that the order of the $j$ and $\xi$ integrations in (\ref{order}) cannot be interchanged in general. This is because in Regge theory the amplitude is obtained by an analytic continuation from the region $t\gg |s|$. } 
Using
 $d^2b= 2\pi zz' \sinh \xi d \xi$, one obtains
  \beq
&& 2\pi z z' \int_{\xi_0}^\infty d\xi \int \frac{dj}{4i}\frac{1-e^{-i\pi j}}{\sin \pi j} \left(\frac{  zz's}
{4\sqrt{\lambda}}\right)^{j-1} \left(\cosh \xi -\frac{z}{z'} - \frac{z'}{z} \right)
\frac{-iMe^{-M_j\xi}}{4\pi R zz'}  \nonumber \\
&&\approx   \frac{-iM}{4R}\sqrt{\frac{\pi \sqrt{\lambda}}{2\tau^3}} \frac{1-e^{-i\pi j_O}}{\sin \pi j_O} e^{(j_O - 1)\tau} 
 \int_{\xi_0}^\infty  d\xi \left(\cosh \xi -\frac{z}{z'} - \frac{z'}{z} \right) \, 
\xi e^{-\frac{\sqrt{\lambda} \xi^2}{2\tau}}\,, \label{order}
\eeq
where  $\xi_0 = |\ln z/z'|$ and $\frac{zz's}{4\sqrt{\lambda}}\equiv e^\tau$. 
Here we have deformed the contour so as to surround 
the branch cut beginning at $j=j_O$, and evaluated the $j$--integral using the saddle point approximation which is valid when $\tau/\sqrt{\lambda}$ is large. The $\xi$ integral then gives
\beq
 \frac{-iM}{8R}\sqrt{\frac{\pi }{2\tau \sqrt{\lambda}}} \frac{1-e^{-i\pi j_O}}{\sin \pi j_O} 
e^{(j_O - 1)\tau} \left( e^{\frac{\tau}{2\sqrt{\lambda}}}\int_{\xi_0-\frac{\tau}
{\sqrt{\lambda}}}^{\xi_0+\frac{\tau}{\sqrt{\lambda}}} d\xi\,  e^{-\frac{\sqrt{\lambda} \xi^2}
{2\tau}} - \left( \frac{z}{z'} +\frac{z'}{z} \right) \, e^{-\frac{\sqrt{\lambda} \xi_0^2}
{2\tau}}\right)\,. \nonumber \\  \label{result}
\eeq
When $\tau/\sqrt{\lambda}$ is large, the first term in the brackets dominates.  Noting also that typically  $\xi_0$ 
should be small since we are scattering objects of the same size, the limits of the 
Gaussian integral can be extended to $\pm \infty$ and we obtain
  \beq
   \frac{-iM\pi }{8R\sqrt{\lambda}} \frac{1-e^{-i\pi j_O}}{\sin \pi j_O}\left(\frac{zz' s}
{4\sqrt{\lambda}}\right)^{\alpha_O(0)-1}\,,
\eeq
 where 
 \beq
 \alpha_O(0)=j_O+\frac{1}{2\sqrt{\lambda}}=1-\frac{M^2-1}{2\sqrt{\lambda}}\,. 
 \label{shift}
 \eeq
We now see explicitly that the intercept 
has shifted by a small amount after the $b$--integration---a possibility we pointed out at the end of 
section \ref{oddoperator}. 
This is due to the $v\sim b^2$ factor in $T^{(1)}$ which modifies the large--$b$ behavior of the amplitude.  Such a shift would not occur if one 
had exchanged the transverse component of the B--field $B_{\pm \perp}$.  
  A curious  consequence of (\ref{shift}) is that the $M_1=k=1$ mode has $\alpha_O(0)=1$, the value commonly associated with the phenomenological  Odderon. At the moment we do not have a deep understanding of this coincidence, especially in terms of the corresponding operator (\ref{SYMoperator1}).
 
 
 Essentially the same shift occurs in the second term of (\ref{get}). After some algebra, one 
finds that the net effect of the second term is simply to replace the prefactor 
$1-\frac{1}{M^2}$ with $1+\frac{1}{M^2}$.
We thus find,
 \beq
 i\int d^2b \, {\mathcal A^{\pm}}(s,b)&\approx &\pm i\left(\frac{N_c}{2\pi \alpha'\mbox{Vol}_{S^4} }
\right)^2 \sum_{I,k} \frac{M_I}{k+2}\left(1+\frac{1}{M_I^2}\right)\int dz d\Omega_4 Y^{(k)}(\Omega_5) \nonumber \\  &&\times
  \int dz'd\Omega'_4 Y^{(k)}(\Omega'_5) \frac{ \kappa^2}{8R^4\pi^2\sqrt{\lambda}}
\frac{1-e^{-i\pi j_O}}{\sin \pi j_O} \left(\frac{zz' s}
{4\sqrt{\lambda}}\right)^{\alpha_O(0)-1} \nonumber \\
  &=&\pm \frac{i\sqrt{\lambda}\pi }{8(\mbox{Vol}_{S^4})^2 } \sum_{I,k}\frac{M_I+\frac{1}{M_I}}{k+2}\int dz d\Omega_4 Y^{(k)}
(\Omega_5) \nonumber \\ && \times
  \int dz'd\Omega'_4 Y^{(k)}(\Omega'_5)  \frac{1-e^{-i\pi j_O}}{\sin \pi j_O}
\left(\frac{zz' s}{4\sqrt{\lambda}}\right)^{\alpha_O(0)-1}
  \,.
 \eeq
The difference between the total cross sections is then given by 
\beq
\Delta \sigma&=&\sigma^{B\bar{B}}-\sigma^{BB}=   2\int d^2b \,\emph{Im}\,{\mathcal A^{-}}(s,b)-2
\int d^2b\, \emph{Im}\, {\mathcal A^{+}}(s,b) \nonumber \\ &=&-\frac{\pi \sqrt{\lambda}}
{4(\mbox{Vol}_{S^4})^2 } \sum_{I,k}\frac{M_I+\frac{1}{M_I}}{k+2} \int dz d\Omega_4 Y^{(k)}(\Omega_5)
  \int dz'd\Omega'_4 Y^{(k)}(\Omega'_5)  \left(\frac{zz' s}{4\sqrt{\lambda}}\right)^{\alpha_O(0)-1}\,. 
  \nonumber \label{fin}\\
\eeq
This is the main result of this paper. 
We remind the reader  that the $z$--integrals are bounded $z\le r_h$, see \eqref{bps}.

\section{Discussion}

Surprisingly, the right hand side of (\ref{fin}) is {\it negative}, which means that the 
baryon--baryon cross section is {\it larger} than the baryon--anti-baryon cross section.  This 
immediately raises both theoretical and practical questions.

Theoretically, one would expect the exchange of an antisymmetric field to generate a repulsive 
force between like charges, as in the well--known cancellation of the attractive NS-NS and repulsive 
R-R forces between parallel D--branes. However, in the above calculation the B--field exchange 
effectively generates an attraction between like charges and repulsion between opposite charges. The 
sign  change can be traced back to the first term in equation \eqref{11} where 
the $z$ and $z'$ derivatives both act on the {\it denominator} of the chordal distance $u$, 
resulting in the first two terms of \eqref{first}. These are positive and 
dominate over the (expected) negative contributions because of the $b^2$ factor which is amplified by the $b$--integration.
Thus the attraction originates from a combined effect of the warp factor of  $AdS_5$ and the 
particular component $B_{\pm z}$ we have used. Had we exchanged the transverse component
 $B_{\pm \perp}$, we would have obtained an opposite sign in the final result.

Practically, the available experimental data show that the total cross section is larger in $p\bar{p}$ collisions  
than in $pp$ collisions. However, before comparing this fact with our result, the 
following three remarks are in order:\\
 (i) The dominance of the first term over the second term in (\ref{result}) is safely 
claimed only at very high energies, $\tau \gg \sqrt{\lambda}$, where the small shift $1/2\sqrt{\lambda}$ in the intercept
 becomes noticeable. On the other hand the highest energy at which $\Delta\sigma$ 
has been measured is  the ISR energy $\sqrt{s}= 53$ GeV which is not that high. 
We would probably need to at least 
go to the Tevatron $\sqrt{s}=2$ TeV, or the LHC energies $\sqrt{s}=14$ TeV. 
\\
(ii) Our result pertains to  a particular choice of the D5--brane embedding which has no dependence on $S^4$. This means 
that the corresponding baryon is a singlet under the $SO(5)$ subgroup of the `flavor' $SO(6)\cong SU(4)$ 
group. As we have seen, this has resulted in the decoupling of  the vector meson, or the Reggeon contribution which plays an important role in QCD.  
 By considering baryons which belong to a larger representation of the flavor group, one should be able 
to find nonvanishing coupling to the Reggeon. In practice, this amounts to adding electric fluxes in 
the $S^4$ direction. The point is that, by a simple generalization of the argument in a previous 
paragraph, one is guaranteed that the exchange of the vector mode $b_\pm$  gives a normal sign. 
 Therefore, in more realistic situations where both the Odderon and the Reggeon couple, the sign of $\Delta\sigma$ is determined by  a competition between the two exchanges. The Reggeon has the intercepts\footnote{It is easy to see
 that the $b$--integral will not modify the intercept in the Reggeon case or in the transverse case 
$B_{\pm \perp}$ due to the absence of the $v\sim b^2$ factor in the tensorial part of the propagator.} 
\beq
\alpha_R(0)=1-\frac{9}{2\sqrt{\lambda}}, \ \ 1-\frac{16}{2\sqrt{\lambda}}, \cdots\,,
\label{ent1}
\eeq
and gives positive contributions to $\Delta\sigma$, whereas the Odderon has the intercepts
\beq 
 \alpha_O(0)=1, \ \  1-\frac{3}{2\sqrt{\lambda}},\ \ 1-\frac{8}{2\sqrt{\lambda}},\ \  1-\frac{15}{2\sqrt{\lambda}}, \cdots\,. \label{ent2}
 \eeq
 and gives negative contributions at very high energies. Which effect wins is likely to become a quantitative question, rather than parametric. \\
 (iii) We have not included unitarity corrections in our calculation. In ${\mathcal N}=4$ SYM proper, strong unitarity corrections come from the eikonal exponentiation of the graviton (Pomeron) amplitude which is dominantly real  \cite{Cornalba:2007zb,Brower:2007qh}, and this could make any `predictions' of the single Odderon approximation  uncertain \cite{Brower:2008cy}. However, as far as unitarity corrections are concerned, ${\mathcal N}=4$ SYM fares poorly as a model of QCD where the Pomeron amplitude is dominantly imaginary. In contrast, the Odderon amplitude is dominantly real both in QCD and ${\mathcal N}=4$ SYM, and in this sense the Odderon sector of the AdS/CFT correspondence is closer to QCD than the Pomeron sector. Turning to phenomenology, in practice there is no urgent need for unitarity corrections to the Odderon, since its intercept $\alpha_O\le 1$ does not violate any constraints from unitarity. Indeed, most of the recent phenomenological applications \cite{Gauron:1986nk,Bartels:2001hw,Dosch:2002ai,Hagler:2002nh,Bartels:2003zu,Bzdak:2007cz,Avila:2006wy,Pire:2008xe,Merino:2009nu} are more or less based on models inspired by the single Odderon exchange.\footnote{In perturbative QCD, unitarity corrections to the BLV Odderon have been included only in asymmetric collisions where one of the hadrons is small and perturbative  \cite{Kovchegov:2003dm,Hatta:2005as,Motyka:2005ep}. } In view of these circumstances, we retain the hope that the qualitative features of the single Odderon exchange can survive  and give useful information to QCD. 
  
  Backed by these observations, we now come to the implications of our results to experiments. The proton belongs to the $\textbf{8}$ 
of the flavor $SU(3)$ group which is not a singlet under any 
subgroup of $SU(3)$. Therefore, it should couple to  both the Reggeon and the Odderon. 
 Then we find the following scenario rather compelling: 
 The ISR data show that $\Delta\sigma>0$ in  
the hitherto explored energy regime. This is quite naturally attributed to the $\alpha_R(0)=1-9/2\sqrt{\lambda}$ Reggeon, 
whereas in the Odderon sector there must be some cancellation 
of the sort mentioned above.  However, the Reggeon contribution $s^{\alpha_R(0)-1}$ dies away quickly as the energy 
increases. On the other hand, we see that certain components of the Odderon have much milder energy dependences. Then 
the Odderon contribution inevitably takes over and $\Delta \sigma$ must turn negative.   
As mentioned in the introduction, the possiblity of having $\Delta \sigma <0$ was 
already raised in \cite{Lukaszuk:1973nt}, and more recently also in \cite{Avila:2006wy} on the basis of an 
extrapolation of a purely phenomenological fit. It is quite remarkable that the AdS/CFT correspondence 
allows for an {\it analytical} evaluation of the cross section difference which leads to the same conclusion.  Incidentally, 
it is amusing to notice that, if one uses $\sqrt{\lambda}=7\sim 8$ for the `t Hooft coupling which 
in QCD ($N_c=3$) would correspond to a typical strong coupling regime  $\alpha_s=1\sim 2$, 
 one obtains $\alpha^{QCD}_R(0) = 0.4\sim 0.5$ in rough accordance with the known phenomenological value. 
Note also that at the ISR energy  the same estimate gives $\tau/\sqrt{\lambda} \approx 1$, suggesting that the Odderon 
contribution is not yet dominant in this regime.
 
 Admittedly, the above scenario sidesteps many perils in naively translating the results for ${\mathcal N}=4$ SYM to those 
for QCD. However, to the extent that we believe in the potential of the AdS/CFT correspondence to shed light on the 
otherwise inaccessible regime of gauge theories, 
   we propose it as a very interesting and testable possibility that carries an imprint of string theory in a curved background. 

Unfortunately, experiments of {\it both} $pp$ and $p\bar{p}$ 
collisions at similar, and high energies (beyond ISR) are lacking. 
$p\bar{p}$ collisions are currently not planned in the ongoing  
program at the LHC, while there are no $pp$ data at the Tevatron. We hope this situation will change in the future, as we expect that the question 
on the fate of the Odderon will most likely be possible to 
answer only when data from new experiments are available.  One possibility 
is the planned $pp$ collisions at RHIC at around $\sqrt{s}=500$ GeV which,  though 
not at very high energy, would still be welcome.

There are several directions for further study. As suggested already, one can employ more realistic baryon configurations 
(see, e.g., \cite{Callan:1999zf}) which have  fluxes in various directions and/or a nontrivial 
extension in the transverse direction. Then the modes $b_\pm$ and $B_{\pm \perp}$ will come into play. 
Moreover, the calculation should be extended to observables other than the total cross section difference. 
There are a number of processes where the Odderon, or more generally, the $C$--odd exchanges are involved 
\cite{Bartels:2001hw,Hagler:2002nh,Bartels:2003zu,Dosch:2002ai,Bzdak:2007cz,Avila:2006wy,Pire:2008xe,Merino:2009nu}, many  of which concern the real part of the amplitude.  It would be interesting to revisit
 these studies with inputs from the AdS/CFT correspondence.  Finally, 
 one would like to understand the difference in the total cross sections in terms of the final states, 
especially in the Odderon exchange channel where the relevant particle production mechanism must be such that it gives a negative contribution to $\Delta \sigma$.  [See, e.g., \cite{Kharzeev:1996sq,Abramovsky:2009ni} 
for studies of final states with emphasis on the difference between $pp$ and $p\bar{p}$ collisions.]    
This might be difficult to test  in the total 
cross section measurement since the Odderon contribution is a small fraction, but perhaps there are ways to see it in less inclusive reactions.  

\section*{Acknowledgments}
We thank Koji Hashimoto, Yoshifumi Hyakutake, Dima Kharzeev, Shin Nakamura, Makoto Sakaguchi,  Yuji Satoh, Anna Stasto and Chung-I Tan for discussions and  
helpful conversations.
E.~A. is supported under the U.S. D.O.E. 
grant number DE-FG02-90-ER-40577. 
 Y.~H.  is supported by Special Coordination Funds of the Ministry of Education, Culture, 
Sports, Science and Technology, the Japanese Government. 

\appendix

\section{Spherical harmonics on $S^5$}

In this appendix we review the basics of the scalar 
and vector spherical harmonics which appear in the decomposition \eqref{ex}. (See also \cite{Lee:1998bxa}.)

\subsection{Scalar spherical harmonics}

We  parameterize the coordinates of $S^5 \in {\mathbb R}^{6}$ as
\beq
y^1&=&\cos \theta\,, \nonumber \\
y^2&=&\sin \theta \cos \theta_2\,, \nonumber \\
y^3&=&\sin \theta \sin \theta_2 \cos \theta_3\,,\nonumber \\
y^4&=&\sin \theta \sin \theta_2 \sin \theta_3 \cos \theta_4\,, \nonumber \\
y^{5}&=&\sin \theta \sin \theta_2 \sin \theta_3\sin \theta_4\cos \theta_5\,, \nonumber \\
y^{6}&=& \sin \theta \sin \theta_2\sin \theta_3\sin \theta_{4}\sin \theta_5\,.
\eeq
The volume element is
\beq
d\Omega_5 = \sin^{4} \theta \sin^{3}\theta_2\sin^2\theta_3 \sin \theta_4 d\theta d\theta_2 d\theta_3 d\theta_{4}d\theta_5 =  \sin^{4}\theta  d\theta d\Omega_{4}\,.
\eeq
 The rank--$k$ scalar spherical harmonics are defined as
 \beq
 Y^{(k)}=\sum C_{i_1,i_2,..i_k} y^{i_1}y^{i_2}\cdots y^{i_k}
 \eeq
 where indices $i$ runs from 1 to $6$ and the tensor $C$ is totally symmetric and traceless with respect to any pair of indices.  
 They form a representation of $SO(6)$. Using the Dynkin label of $SU(4)\cong SO(6)$, it is 
 \beq
 (0,k,0)\,,
 \eeq
 whose dimension is
 \beq
 d=\frac{(k+3)(k+2)^2 (k+1)}{12}\,.
 \eeq
 Because of the traceless condition, one has that
 \beq
 \int_{S^5} Y^{(k)} =0\,.
 \eeq
 For $k=1$, there are six harmonics  which are just the six coordinates $y^1,y^2,..,y^{6}$.
Note that $Y^{(1)}\propto y^1=\cos \theta$ is independent of the $S^{4}$ coordinates, so it 
 gives a nonzero value after integration over $S^4$ as in (\ref{couple}).

 For $k=2$, the allowed tensor is, for example,
 \beq
 C_{ij}=\delta_{i1}\delta_{j2}+\delta_{i2}\delta_{j1}\,,
 \eeq
 which gives
 \beq  Y^{(2)}\propto y^1y^2\,.
 \eeq
 This vanishes upon integration over $S^{4}$.
 Another possibility is
 \beq
 C_{ij}=\delta_{i1}\delta_{j1}-\frac{1}{6}\delta_{ij}\,,
 \eeq
  so that 
 \beq
 Y^{(2)}\propto \cos^2\theta-\frac{1}{6}\,.
 \eeq
 This is nonzero after integrating over $S^{4}$. For any value of $k$ there is at least one spherical harmonics
  which does not vanish after integrated over $S^{4}$.

\subsection{Vector spherical harmonics}

The rank-$k$ vector spherical harmonics are defined by
\beq
Y^{(k)}_i=\sum C^i_{i_1,i_2,..i_k} y^{i_1}y^{i_2}\cdots y^{i_k} \label{dir}
 \eeq
 with the conditions that
 \beq
 \partial_i Y_i= y_i Y_i=0\,. \label{cond}
 \eeq
Projecting on $S^5$, one finds
\beq Y_\alpha^{(k)}=\frac{\partial y^i}{\partial \theta^\alpha}Y_i^{(k)}\,.
\eeq
Useful identities are
\beq
\nabla^\alpha Y^{(k)}_\alpha=0\,, \qquad \nabla^2  Y_\alpha^{(k)}=-\bigl(k(k+4)-1\bigr)Y_\alpha^{(k)}\,.
\eeq

 One sees that a vector of the form (\ref{dir}) is a direct product of the fundamental representation of $SO(6)$ and the totally symmetric rank--$k$ tensor. Using the Dynkin label, one has the decomposition
\beq
(0,1,0)\otimes (0,k,0)=(0,k+1,0)\oplus (1,k-1,1)\oplus (0,k-1,0)\,.
\eeq
$(1,k-1,1)$ is the rank--$k$ vector spherical harmonics whose dimension is
\beq
d=\frac{k(k+2)^2(k+4)}{3}\,.
\eeq
For $k=1$, $d=15$ which is the number of the Killing vectors.
The conditions (\ref{cond}) become
 \beq
 y^iC^i_jy^j=C^i_i=0\,.
 \eeq
 The solution is, for example,
 \beq
 C^i_j=\delta_{i1}\delta_{j2}-\delta_{i2}\delta_{j1}\,,
 \eeq
 so that
 \beq
 Y_\alpha^{(1)}\propto y_2\partial_\alpha y_1-y_1\partial_\alpha y_2\,.
 \eeq
 This is indeed a Killing vector. It is easy to show that
 \beq
 \int_{S^4}Y^{(1)}_\theta=0\,,
 \eeq
 for all the 15 Killing vectors.

For $k=2$, we have 64 harmonics which appears in the decomposition
\beq
6\otimes 20 = 50 \oplus 64 \oplus 6\,.
\eeq
 They can be constructed as follows. Starting from an arbitrary tensor $A^i_{jk}$ which is symmetric and traceless in $j$ and $k$, one has the decomposition
\beq
A^i_{jk}&=&\frac{1}{3}\left(A^i_{jk}+A^j_{ik}+A^k_{ij}-\frac{\delta_{ij}}{4}A^l_{lk}-\frac{\delta_{ik}}{4}A^l_{lj}
-\frac{\delta_{jk}}{4}
A^l_{li} \right)\nonumber \\
&+& \frac{1}{3}\left(2A^i_{jk}-A^j_{ik}-A^k_{ij}\right) -\frac{1}{15}\left(\delta_{ij}A^l_{lk}+\delta_{ik}A^l_{lj}-2\delta_{jk}A^l_{li}\right)\nonumber \\
&+& \frac{1}{20}\left(3\delta_{ij}A^l_{lk}+3\delta_{ik}A^l_{lj}-\delta_{jk}A^l_{li} \right)\,.
\eeq
The first line is totally symmetric and traceless in $ijk$, so this belongs to the $k=3$ scalar spherical harmonics $(0,3,0)$. Projecting on $S^5$, it becomes $\nabla_\alpha Y^{(3)}$.  The last line is $(0,1,0)$ which becomes $\nabla_\alpha Y^{(1)}$ on $S^5$.

The second line is the $k=2$ vector spherical harmonics
\beq
C^i_{jk}=\frac{1}{3}\left(2A^i_{jk}-A^j_{ik}-A^k_{ij}\right) -\frac{1}{15}\left(\delta_{ij}A^l_{lk}+\delta_{ik}A^l_{lj}-2\delta_{jk}A^l_{li}\right)\,, \label{a}
\eeq
constructed such that it satisfies (\ref{cond}) which reads
\beq
C^i_{jk}y^iy^jy^k=0, \qquad C^i_{ij}y^j=0\,. \label{co}
\eeq

We are now ready to prove that
\beq
\int_{S^4} Y_{\theta}^{(2)}=0\,.
\eeq
Using the notation $y^a=y^{2,3,4,5,6}$, we have
\beq
Y_{\theta}^{(2)}&=&\frac{\partial y^i}{\partial \theta} C^i_{jk} y^jy^k=\frac{\partial y^1}{\partial \theta} C^1_{jk} y^jy^k+\frac{\partial y^a}{\partial \theta} C^a_{jk} y^jy^k  \nonumber \\
&=&-\sin \theta  C^1_{jk}
y^jy^k+\frac{\cos \theta}{\sin \theta} y^a C^a_{jk} y^jy^k =\left( -\sin \theta-\frac{\cos^2 \theta}{\sin \theta}\right) C^1_{jk}y^jy^k\,,
\eeq
where in the last equality we used the first condition of (\ref{co}).
The last term can be written as
\beq
C^1_{jk}y^jy^k =\cos^2 \theta C^1_{11} + 2\cos \theta \sin \theta C^1_{1a} \hat{y}^a + \sin^2 \theta  C^1_{ab}\hat{y}^a \hat{y}^b\,,
\eeq
 where we have denoted $y^a=\sin \theta \, \hat{y}^a$.
The first term is zero because $C^1_{11}=0$
as is evident from (\ref{a}).
The second term is $k=1$ scalar spherical harmonics on $S^4$, so it vanishes after integrating over $S^4$. The last term gives, after $S^4$ integration,
\beq
\int C^1_{ab}\hat{y}^a \hat{y}^b \propto C^1_{ab}\delta_{ab}=C^1_{aa}\,.
\eeq
However, this is zero because of the traceless condition $C^1_{jj}=C^1_{11}+C^1_{aa}=0$.
Thus we have proven that the $k=2$ vector spherical harmonics do not couple to the D--brane.

For the $k=3$ vector spherical harmonics, we have similarly,
\beq
Y^{(3)}_\theta \sim C^1_{111}\cos^3 \theta + 3\cos^2 \theta \sin \theta C^1_{11a}\hat{y}^a + 3\cos \theta \sin^2 \theta
C^1_{1ab}\hat{y}^a\hat{y}^b + \sin^3 \theta C^1_{abc}\hat{y}^a\hat{y}^b\hat{y}^c\,. \nonumber 
\eeq
The second and the fourth terms give zero after integrating over $S^4$. The third term gives
$C^1_{1aa}\,$. 
This is zero if $C^1_{111}=0$ (because of the traceless condition $C^1_{1jj}=0$), in which case the first term also vanishes. To see this is indeed the case, note that  (\ref{cond}) reads
\beq
0=C^i_{jkl}y^iy^jy^ky^l=\cos^4 \theta C^1_{111} +\cos^3\theta \sin \theta(3C^1_{11a}+C^a_{111})\hat{y}^a +\cdots\,.
\eeq
 In order for this to hold identically, one must have that $C^1_{111}=0$, $3C^1_{11a}+C^a_{111}=0$, and so on.

 The situation is similar for higher $k$ values. First one has $C^1_{1111\cdots}=0$. After integrating over $S^4$, one gets terms like
 \beq
 C^1_{1aabbcc\dots}\,, \quad C^1_{111aabb\dots}\,,
 \eeq
 for $k$ odd, and
 \beq
 C^1_{aabbcc\dots}\,, \quad C^1_{11aabb\dots}\,,
 \eeq
  for $k$ even. These are all zero because, for instance,
  \beq
  C^1_{1aabbcc}=-C^1_{111bbcc}=C^1_{11111cc}=-C^1_{1111111}=0\,.
  \eeq
 Thus we have proven that the integral (\ref{prove}) vanishes for all $k$.


\bibliographystyle{JHEP}
\bibliography{odderon}

\providecommand{\href}[2]{#2}\begingroup\raggedright\begin{thebibliography}{10}

\bibitem{Lukaszuk:1973nt}
L.~Lukaszuk and B.~Nicolescu, {\it {A Possible interpretation of p p rising
  total cross- sections}},  {\em Nuovo Cim. Lett.} {\bf 8} (1973) 405--413.

\bibitem{Ewerz:2003xi}
C.~Ewerz, {\it {The Odderon in quantum chromodynamics}},
  \href{http://xxx.lanl.gov/abs/hep-ph/0306137}{{\tt hep-ph/0306137}}.

\bibitem{Gauron:1986nk}
P.~Gauron, B.~Nicolescu, and E.~Leader, {\it {Similarities and differences
  between anti-p p and p p scattering at TeV energies and beyond}},  {\em Nucl.
  Phys.} {\bf B299} (1988) 640.

\bibitem{Bartels:2001hw}
J.~Bartels, M.~A. Braun, D.~Colferai, and G.~P. Vacca, {\it {Diffractive
  $\eta_c$ photo- and electroproduction with the perturbative QCD odderon}},
  {\em Eur. Phys. J.} {\bf C20} (2001) 323--331,
  [\href{http://xxx.lanl.gov/abs/hep-ph/0102221}{{\tt hep-ph/0102221}}].

\bibitem{Dosch:2002ai}
H.~G. Dosch, C.~Ewerz, and V.~Schatz, {\it {The odderon in high energy elastic
  p p scattering}},  {\em Eur. Phys. J.} {\bf C24} (2002) 561--571,
  [\href{http://xxx.lanl.gov/abs/hep-ph/0201294}{{\tt hep-ph/0201294}}].

\bibitem{Hagler:2002nh}
P.~Hagler, B.~Pire, L.~Szymanowski, and O.~V. Teryaev, {\it {Hunting the
  QCD-odderon in hard diffractive electroproduction of two pions}},  {\em Phys.
  Lett.} {\bf B535} (2002) 117--126,
  [\href{http://xxx.lanl.gov/abs/hep-ph/0202231}{{\tt hep-ph/0202231}}].

\bibitem{Bartels:2003zu}
J.~Bartels, M.~A. Braun, and G.~P. Vacca, {\it {The process $\gamma$(*) + p
  $\to$ $\eta_c$ + X: A test for the perturbative QCD odderon}},  {\em Eur.
  Phys. J.} {\bf C33} (2004) 511--521,
  [\href{http://xxx.lanl.gov/abs/hep-ph/0304160}{{\tt hep-ph/0304160}}].

\bibitem{Bzdak:2007cz}
A.~Bzdak, L.~Motyka, L.~Szymanowski, and J.~R. Cudell, {\it {Exclusive J/psi
  and Upsilon hadroproduction and the QCD odderon}},  {\em Phys. Rev.} {\bf
  D75} (2007) 094023, [\href{http://xxx.lanl.gov/abs/hep-ph/0702134}{{\tt
  hep-ph/0702134}}].

\bibitem{Avila:2006wy}
R.~Avila, P.~Gauron, and B.~Nicolescu, {\it {How can the Odderon be detected at
  RHIC and LHC}},  {\em Eur. Phys. J.} {\bf C49} (2007) 581--592,
  [\href{http://xxx.lanl.gov/abs/hep-ph/0607089}{{\tt hep-ph/0607089}}].

\bibitem{Pire:2008xe}
B.~Pire, F.~Schwennsen, L.~Szymanowski, and S.~Wallon, {\it {Hard
  Pomeron-Odderon interference effects in the production of $\pi^+\pi^-$ pairs
  in high energy gamma-gamma collisions at the LHC}},  {\em Phys. Rev.} {\bf
  D78} (2008) 094009, [\href{http://xxx.lanl.gov/abs/0810.3817}{{\tt
  arXiv:0810.3817}}].

\bibitem{Merino:2009nu}
C.~Merino, M.~M. Ryzhinskiy, and Y.~M. Shabelski, {\it {Odderon Effects in pp
  Collisions: Predictions for LHC Energies}},
  \href{http://xxx.lanl.gov/abs/0906.2659}{{\tt arXiv:0906.2659}}.

\bibitem{Domokos:2009hm}
S.~K. Domokos, J.~A. Harvey, and N.~Mann, {\it {The Pomeron contribution to p p
  and p bar p scattering in AdS/QCD}},
  \href{http://xxx.lanl.gov/abs/0907.1084}{{\tt arXiv:0907.1084}}.

\bibitem{Donnachie:1992ny}
A.~Donnachie and P.~V. Landshoff, {\it {Total cross-sections}},  {\em Phys.
  Lett.} {\bf B296} (1992) 227--232,
  [\href{http://xxx.lanl.gov/abs/hep-ph/9209205}{{\tt hep-ph/9209205}}].

\bibitem{Brower:2008cy}
R.~C. Brower, M.~Djuric, and C.-I. Tan, {\it {Odderon in gauge/string
  duality}},  {\em JHEP} {\bf 07} (2009) 063,
  [\href{http://xxx.lanl.gov/abs/0812.0354}{{\tt arXiv:0812.0354}}].

\bibitem{Bartels:1980pe}
J.~Bartels, {\it {High-Energy Behavior in a Nonabelian Gauge Theory. 2. First
  Corrections to T(n$\to$ m) Beyond the Leading LNS Approximation}},  {\em
  Nucl. Phys.} {\bf B175} (1980) 365.

\bibitem{Kwiecinski:1980wb}
J.~Kwiecinski and M.~Praszalowicz, {\it {Three Gluon Integral Equation and Odd
  c Singlet Regge Singularities in QCD}},  {\em Phys. Lett.} {\bf B94} (1980)
  413.

\bibitem{Lipatov:1993yb}
L.~N. Lipatov, {\it {High-energy asymptotics of multicolor QCD and exactly
  solvable lattice models}},
  \href{http://xxx.lanl.gov/abs/hep-th/9311037}{{\tt hep-th/9311037}}.

\bibitem{Faddeev:1994zg}
L.~D. Faddeev and G.~P. Korchemsky, {\it {High-energy QCD as a completely
  integrable model}},  {\em Phys. Lett.} {\bf B342} (1995) 311--322,
  [\href{http://xxx.lanl.gov/abs/hep-th/9404173}{{\tt hep-th/9404173}}].

\bibitem{Janik:1998xj}
R.~A. Janik and J.~Wosiek, {\it {Solution of the odderon problem}},  {\em Phys.
  Rev. Lett.} {\bf 82} (1999) 1092--1095,
  [\href{http://xxx.lanl.gov/abs/hep-th/9802100}{{\tt hep-th/9802100}}].

\bibitem{Kotanski:2006ec}
J.~Kotanski, {\it {Three particle Pomeron and odderon states in QCD}},  {\em
  Acta Phys. Polon.} {\bf B37} (2006) 2615--2654,
  [\href{http://xxx.lanl.gov/abs/hep-th/0603238}{{\tt hep-th/0603238}}].

\bibitem{Bartels:1999yt}
J.~Bartels, L.~N. Lipatov, and G.~P. Vacca, {\it {A New Odderon Solution in
  Perturbative QCD}},  {\em Phys. Lett.} {\bf B477} (2000) 178--186,
  [\href{http://xxx.lanl.gov/abs/hep-ph/9912423}{{\tt hep-ph/9912423}}].

\bibitem{Hatta:2005as}
Y.~Hatta, E.~Iancu, K.~Itakura, and L.~McLerran, {\it {Odderon in the color
  glass condensate}},  {\em Nucl. Phys.} {\bf A760} (2005) 172--207,
  [\href{http://xxx.lanl.gov/abs/hep-ph/0501171}{{\tt hep-ph/0501171}}].

\bibitem{Kovchegov:2003dm}
Y.~V. Kovchegov, L.~Szymanowski, and S.~Wallon, {\it {Perturbative odderon in
  the dipole model}},  {\em Phys. Lett.} {\bf B586} (2004) 267--281,
  [\href{http://xxx.lanl.gov/abs/hep-ph/0309281}{{\tt hep-ph/0309281}}].

\bibitem{Braun:2007kz}
M.~A. Braun, {\it {Odderon with a running coupling constant}},  {\em Eur. Phys.
  J.} {\bf C53} (2008) 59--63, [\href{http://xxx.lanl.gov/abs/0707.2314}{{\tt
  arXiv:0707.2314}}].

\bibitem{Stasto:2009bc}
A.~M. Stasto, {\it {Small x resummation and the Odderon}},  {\em Phys. Lett.}
  {\bf B679} (2009) 288--292, [\href{http://xxx.lanl.gov/abs/0904.4124}{{\tt
  arXiv:0904.4124}}].

\bibitem{Braunewell:2005ct}
S.~Braunewell and C.~Ewerz, {\it {The C-odd four-gluon state in the color glass
  condensate}},  {\em Nucl. Phys.} {\bf A760} (2005) 141--171,
  [\href{http://xxx.lanl.gov/abs/hep-ph/0501110}{{\tt hep-ph/0501110}}].

\bibitem{Kovner:2005qj}
A.~Kovner and M.~Lublinsky, {\it {Odderon and seven Pomerons: QCD Reggeon field
  theory from JIMWLK evolution}},  {\em JHEP} {\bf 02} (2007) 058,
  [\href{http://xxx.lanl.gov/abs/hep-ph/0512316}{{\tt hep-ph/0512316}}].

\bibitem{Motyka:2005ep}
L.~Motyka, {\it {Nonlinear evolution of pomeron and odderon in momentum
  space}},  {\em Phys. Lett.} {\bf B637} (2006) 185--191,
  [\href{http://xxx.lanl.gov/abs/hep-ph/0509270}{{\tt hep-ph/0509270}}].

\bibitem{Jeon:2005cf}
S.~Jeon and R.~Venugopalan, {\it {A classical odderon in QCD at high
  energies}},  {\em Phys. Rev.} {\bf D71} (2005) 125003,
  [\href{http://xxx.lanl.gov/abs/hep-ph/0503219}{{\tt hep-ph/0503219}}].

\bibitem{Janik:1999zk}
R.~A. Janik and R.~B. Peschanski, {\it {High energy scattering and the AdS/CFT
  correspondence}},  {\em Nucl. Phys.} {\bf B565} (2000) 193--209,
  [\href{http://xxx.lanl.gov/abs/hep-th/9907177}{{\tt hep-th/9907177}}].

\bibitem{Kim:1985ez}
H.~J. Kim, L.~J. Romans, and P.~van Nieuwenhuizen, {\it {The Mass Spectrum of
  Chiral N=2 D=10 Supergravity on S**5}},  {\em Phys. Rev.} {\bf D32} (1985)
  389.

\bibitem{Brower:2006ea}
R.~C. Brower, J.~Polchinski, M.~J. Strassler, and C.-I. Tan, {\it {The Pomeron
  and Gauge/String Duality}},  {\em JHEP} {\bf 12} (2007) 005,
  [\href{http://xxx.lanl.gov/abs/hep-th/0603115}{{\tt hep-th/0603115}}].

\bibitem{Hatta:2009ra}
Y.~Hatta, T.~Ueda, and B.-W. Xiao, {\it {Polarized DIS in N=4 SYM: Where is
  spin at strong coupling?}},  {\em JHEP} {\bf 08} (2009) 007,
  [\href{http://xxx.lanl.gov/abs/0905.2493}{{\tt arXiv:0905.2493}}].

\bibitem{D'Hoker:2002aw}
E.~D'Hoker and D.~Z. Freedman, {\it {Supersymmetric gauge theories and the
  AdS/CFT correspondence}},  \href{http://xxx.lanl.gov/abs/hep-th/0201253}{{\tt
  hep-th/0201253}}.

\bibitem{Bachas:1995kx}
C.~Bachas, {\it {D-brane dynamics}},  {\em Phys. Lett.} {\bf B374} (1996)
  37--42, [\href{http://xxx.lanl.gov/abs/hep-th/9511043}{{\tt
  hep-th/9511043}}].

\bibitem{Cornalba:2007zb}
L.~Cornalba, M.~S. Costa, and J.~Penedones, {\it {Eikonal Approximation in
  AdS/CFT: Resumming the Gravitational Loop Expansion}},  {\em JHEP} {\bf 09}
  (2007) 037, [\href{http://xxx.lanl.gov/abs/0707.0120}{{\tt
  arXiv:0707.0120}}].

\bibitem{Brower:2007qh}
R.~C. Brower, M.~J. Strassler, and C.-I. Tan, {\it {On the Eikonal
  Approximation in AdS Space}},  {\em JHEP} {\bf 03} (2009) 050,
  [\href{http://xxx.lanl.gov/abs/0707.2408}{{\tt arXiv:0707.2408}}].

\bibitem{Imamura:1998gk}
Y.~Imamura, {\it {Supersymmetries and BPS configurations on Anti-de Sitter
  space}},  {\em Nucl. Phys.} {\bf B537} (1999) 184--202,
  [\href{http://xxx.lanl.gov/abs/hep-th/9807179}{{\tt hep-th/9807179}}].

\bibitem{Callan:1998iq}
J.~Callan, Curtis~G., A.~Guijosa, and K.~G. Savvidy, {\it {Baryons and string
  creation from the fivebrane worldvolume action}},  {\em Nucl. Phys.} {\bf
  B547} (1999) 127--142, [\href{http://xxx.lanl.gov/abs/hep-th/9810092}{{\tt
  hep-th/9810092}}].

\bibitem{Witten:1998xy}
E.~Witten, {\it {Baryons and branes in anti de Sitter space}},  {\em JHEP} {\bf
  07} (1998) 006, [\href{http://xxx.lanl.gov/abs/hep-th/9805112}{{\tt
  hep-th/9805112}}].

\bibitem{Arutyunov:1998hf}
G.~E. Arutyunov and S.~A. Frolov, {\it {Quadratic action for type IIB
  supergravity on AdS(5) x S(5)}},  {\em JHEP} {\bf 08} (1999) 024,
  [\href{http://xxx.lanl.gov/abs/hep-th/9811106}{{\tt hep-th/9811106}}].

\bibitem{Bena:2000fp}
I.~Bena, H.~Nastase, and D.~Vaman, {\it {Propagators for p-forms in AdS(2p+1)
  and correlation functions in the AdS(7)/(2,0) CFT correspondence}},  {\em
  Phys. Rev.} {\bf D64} (2001) 106009,
  [\href{http://xxx.lanl.gov/abs/hep-th/0008239}{{\tt hep-th/0008239}}].

\bibitem{Callan:1999zf}
J.~Callan, Curtis~G., A.~Guijosa, K.~G. Savvidy, and O.~Tafjord, {\it {Baryons
  and flux tubes in confining gauge theories from brane actions}},  {\em Nucl.
  Phys.} {\bf B555} (1999) 183--200,
  [\href{http://xxx.lanl.gov/abs/hep-th/9902197}{{\tt hep-th/9902197}}].

\bibitem{Kharzeev:1996sq}
D.~Kharzeev, {\it {Can Gluons Trace Baryon Number?}},  {\em Phys. Lett.} {\bf
  B378} (1996) 238--246, [\href{http://xxx.lanl.gov/abs/nucl-th/9602027}{{\tt
  nucl-th/9602027}}].

\bibitem{Abramovsky:2009ni}
V.~A. Abramovsky and N.~V. Radchenko, {\it {Possible difference between
  multiplicity distributions and inclusive spectra of secondary hadrons in
  proton-proton and proton-antiproton collisions at energy sqrt(s)=900 GeV}},
  \href{http://xxx.lanl.gov/abs/0912.1041}{{\tt arXiv:0912.1041}}.

\bibitem{Lee:1998bxa}
S.~Lee, S.~Minwalla, M.~Rangamani, and N.~Seiberg, {\it {Three-point functions
  of chiral operators in D = 4, N = 4 SYM at large N}},  {\em Adv. Theor. Math.
  Phys.} {\bf 2} (1998) 697--718,
  [\href{http://xxx.lanl.gov/abs/hep-th/9806074}{{\tt hep-th/9806074}}].

\end{thebibliography}\endgroup

\end{document}